\documentclass[a4paper,11pt]{article}
\pdfoutput=1
\usepackage{others/jcappub} 
\usepackage{here}

\usepackage{ulem}

\newcommand{\bm}[1]{\hbox{\boldmath{$#1$}}}
\newcommand{\sbm}[1]{\hbox{\boldmath{\scriptsize$#1$}}}

\title{\boldmath Efficient self-resonance instability from axions}

\author[a]{Hayato Fukunaga,}
\author[a]{Naoya Kitajima,}
\author[a,b]{Yuko Urakawa}

\affiliation[a]{Department of Physics and Astrophysics, Nagoya University, Chikusa, Nagoya 464-8602, Japan}
\affiliation[b]{Department of Astroparticles and Cosmology, Bielefeld
	University, Universit$\ddot{\rm a}$tsstra{\ss}e 25}

\emailAdd{fukunaga.hayato@c.mbox.nagoya-u.ac.jp}
\emailAdd{kitajima.naoya@f.mbox.nagoya-u.ac.jp}
\emailAdd{urakawa.yuko@h.mbox.nagoya-u.ac.jp}

\abstract{It was recently shown that a coherent oscillation of an axion can cause an efficient parametric resonance, leading to a prominent emission of the gravitational waves (GWs). In this paper, conducting the Floquet analysis, we investigate the parametric resonance instability, which potentially triggers the emission of the GWs from axions. Such a resonance instability takes place, when the time evolution of the background field significantly deviates from the harmonic oscillation. Therefore, the resonance instability cannot be described by the Mathieu equation, whose stability/instability chart is well known. In this paper, introducing an explicitly calculable parameter $\tilde{q}$, which can be used to classify different types of the parametric resonance described by the general Hill's equation, we investigate the stability/instability chart for the general Hill's equation. This can also apply to the case where the background oscillation is anharmonic. We show that the flapping resonance instability, which takes place for $\tilde{q}=O(1)$, typically leads to the most significant growth of the inhomogeneous modes among the self-resonance instability. We also investigate whether the flapping resonance takes place for the cosine potential or not.}

\arxivnumber{1903.02119}

\usepackage{amsmath}	
\begin{document}

\maketitle
\flushbottom

\section{Introduction}

\label{sec:intro}
The parametric resonance in cosmology has been vastly studied in the context of the reheating, which is the transition period from inflation to the hot Big Bang Universe~\cite{Kofman:1994rk,Kofman:1997yn,Greene:1997fu,Traschen:1990sw,Dolgov:1989us} (for a recent review, see Ref.~\cite{Amin:2014eta}). During this period, the coherent oscillation of the inflaton, which is initially almost homogeneous, leads to the resonant production of the inflaton {\it particles} or other particle species with finite momenta, resulting in highly inhomogeneous configuration at sub-Hubble scales.  The efficiency of the parametric resonance and the resultant spectrum of the produced particles crucially depend on model parameters. In particular, when the time evolution of the background inflaton is well approximated by the harmonic oscillation, the parametric resonance instability in the linear regime can be described by the Mathieu equation, whose resonance band is well known in the literature~\cite{mclachlan1947theory}. For the $\lambda \phi^4$ theory, the mode equation can be described by the Lam\'{e} equation, whose resonance band structure is also known~\cite{Greene:1997fu}.

More recently, it was shown that the parametric resonance also takes place during the radiation and matter dominated eras in the context of the axion cosmology~\cite{Soda:2017dsu, Kitajima:2018zco} (see also Refs.~\cite{Enqvist:2008be,Sainio:2012rp}, where the parametric resonance in the curvaton scenario was discussed). Axion is a scalar field which is typically long-lived, being a candidate of dark matter during its oscillation~\cite{Preskill:1982cy, Abbott:1982af,Dine:1982ah}. 
In particular, string theory predicts many axions and the Universe filled with such stringy axions is called the string axiverse \cite{Arvanitaki:2009fg}.
Depending on the potential and initial conditions, the axion can result in a rather efficient parametric resonance. (See Ref.~\cite{Greene:1998pb} for an early work.)


As has been studied in the context of the reheating, the parametric resonance instability can lead to a prominent emission of the gravitational waves (GWs)~\cite{Khlebnikov:1997di,Easther:2006gt,Easther:2006vd,GarciaBellido:2007dg,GarciaBellido:2007af, Antusch:2017flz,Antusch:2016con,Adshead:2018doq}, transforming an almost homogeneous initial distribution into a highly inhomogeneous one. Recently, in Refs.~\cite{Soda:2017dsu, Kitajima:2018zco}, it was shown that the axions which start to oscillate during the radiation and matter dominated eras also can be a prominent source of the gravitational wave (GW) background. In general, axions have self-interactions. The parametric resonance instability driven by the self-interaction, which is sometimes called the self-resonance, can lead to a copious emission of GWs. In particular, string theory predicts the presence of axions in a wide mass range~\cite{Arvanitaki:2009fg}, including very light axions~(see e.g., Refs.~\cite{Conlon:2005ki, Hebecker:2018yxs}). As a consequence of the efficient parametric resonance,  stringy axions with different mass scales can lead to the emissions of the GWs with different frequencies, dubbed GW forest~\cite{Kitajima:2018zco}. This is in contrast to the GWs emitted during the reheating, whose frequency is typically very high. The multi-wavelength observations of the GWs from axions may uncover the mass spectrum of the axions, opening a new window of the string axiverse.



In an expanding Universe, the cosmic expansion typically renders the parametric resonance much less efficient than in the flat space. Therefore, one may wonder why the self-interaction of axions can drive the efficient resonance instability in the expanding Universe. The key to understand this is that, in certain cases, the commencement of the axion oscillation can be significantly delayed . In that case, the parametric resonance can be kept efficient without being disturbed by the cosmic expansion. 
Note that, in such cases, the time evolution of the background homogeneous axion significantly deviates from the harmonic oscillation. Then, the mode equation for the linear perturbation of the axion cannot be approximated by the Mathieu equation. Instead, we should consider the equation for a more general (quasi) periodic function such as the Hill's equation. The stability/instability chart for the general Hill's equation is still poorly understood, compared to the one for the Mathieu equation, which is the simplest case of the Hill's equation (for existing studies of the Hill's equation, see Refs.~\cite{Koutvitsky_PLA,Koutvitsky:2005ds,Koutvitsky:2015mda,Koutvitsky:2018afp,Braden:2010wd,Zanchin:1997gf,Adams:2009ioa,Lachapelle:2008sy,Lozanov:2017hjm}). As a consequence of the anharmonic oscillation of the background axion, the  resonance instability which leads to the prominent emission of the GWs can be qualitatively different from the one described by the Mathieu equation. In particular, in Ref.~\cite{Kitajima:2018zco}, two of the authors pointed out a new type of resonance instability, dubbed the {\it flapping} resonance instability.

Along this line, in this paper, we investigate the stability/instability chart of the parametric resonance described by the general Hill's equation in an expanding Universe. 
Generalizing the $q$-parameter in the Mathieu equation, which is the coefficient of the sinusoidal term and characterizes the resonance band structure. We introduce an explicitly calculable parameter $\tilde{q}$ and show that $\tilde{q}$ characterizes the band width and the growth rate for more general periodic system described by the Hill's equation.
%
%
We will find that the flapping resonance instability takes place when $\tilde{q}$ is ${\cal O}(1)$ and it is typically the most efficient resonance among possible self-resonance instability.

This paper is organized as follows. In Sec.~\ref{Sec:Resonance}, after describing our setup of the problem, we briefly summarize the basic properties of the flapping resonance, clarifying the difference from the conventional narrow and broad resonances described by the Mathieu equation. In Sec.~\ref{Sec:Hill}, we study the stability/instability chart for the Hill's equation, by introducing a new parameter $\tilde{q}$, which characterizes the different types of the parametric resonance. In Sec.~\ref{sec:cos}, we investigate whether a sustainable flapping resonance can take place for the cosine potential or not, having an application to axions in mind. In Sec.~\ref{sec:conc}, we conclude and discuss future issues.


\section{Different types of parametric resonance} \label{Sec:Resonance}
In this section, first, we describe the setup of the problem. Then,
after a brief review of the conventional parametric resonance described by the Mathieu equation, we
discuss the basic properties of the flapping resonance instability pointed out recently in Ref.~\cite{Kitajima:2018zco}, clarifying the difference from the conventional parametric resonance.

\subsection{Preliminaries}  \label{SSec:notation} 
In this paper, we investigate the resonance instability of a general scalar field
$\phi$ whose background homogeneous mode undergoes a periodic
oscillation. Without loss of generality, we can choose the field value $\phi$ such that 
oscillates around $\phi=0$. Having an application to an axion in mind (while our argument can apply more generically), we express the scalar potential as 
\begin{equation}
V(\phi)=(mf)^2\, \tilde{V}(\tilde{\phi}),\qquad\qquad\tilde{\phi} \equiv \frac{\phi}{f}\,,
\end{equation}
where the parameters $m$ and $f$ correspond respectively to the mass and the decay
constant in case $\phi$ is an axion. Particularly for axions, we impose the following
conditions on $\tilde{V}$: i) $Z_2$ symmetry, i.e., the symmetry under
$\phi \to - \phi$, ii) $\tilde{V}(\tilde{\phi}) \sim  \tilde{\phi}^2/2$ for $|\tilde{\phi}| \ll 1$. Then, $\phi$ behaves as dark matter, when it oscillates in $|\tilde{\phi}| \ll 1$.
In this parametrization, $m^2$ gives the curvature of the potential around the minimum. 

When the dilute instanton gas approximation holds, the scalar potential for the axion is given by 
\begin{align}
 &  \tilde{V}(\tilde{\phi}) =1-\cos\tilde{\phi} \,. \label{Exp:cosine}
\end{align}
Meanwhile, as was argued, e.g., in Refs.~\cite{Witten:1979vv, Witten:1980sp}, when this
approximation is violated, $\tilde{V}$ can deviate from the conventional
cosine form especially for $|\tilde{\phi}| \geq 1$.  More recently, the scalar potential of an axion which interacts with 
an SU($N$) gauge field in the large $N$ limit was considered in the
context of axion inflation in Refs.~\cite{Yonekura:2014oja, Nomura:2017ehb}, where it was shown that the scalar potential
of the axion (in a single branch) is given by~\cite{Nomura:2017ehb}      
\begin{align}
 & \tilde{V}(\tilde{\phi}) = \frac{1}{2} \left[  1 - \frac{1}{(1 + \tilde{\phi}^2/c)^c} \right] \qquad
 \quad (c > 0)\,,   \label{purenatural}
\end{align} 
where the scalar potential is connected to a plateau region for
$|\tilde{\phi}| > 1$. (The parameters here and those
in Ref.~\cite{Nomura:2017ehb} are related as $F = \sqrt{c} f$ and 
$M^2 =mf/\sqrt{2}$.) This inflation model was dubbed as pure natural inflation~\cite{Nomura:2017ehb}.   
Other examples where the scalar potential of an axion can deviate
from the cosine form were discussed, e.g., in Refs.~\cite{Soda:2017dsu, Kitajima:2018zco}.

When $\tilde{\phi}$ oscillates around $\tilde{\phi}=0$, the resonance
instability, which can convert an almost homogeneous configuration to
a highly inhomogeneous one, takes place. In this paper, assuming that
$\phi$ is a sub-dominant component of the Universe, we solve the time
evolution of $\phi$ in a fixed spacetime geometry. 
For instance, this is the case
for an axion dark matter which starts to oscillate before the
matter-radiation equality. Let us emphasize that several aspects which
will be discussed in this paper hold as well as in case $\phi$ is the dominant component of the Universe at the onset of the oscillation.

\subsection{Field equation}\label{sec:fieldeq}
In the spatially flat FLRW Universe, the evolution equation for the scalar field ($\tilde{\phi}$) is given by 
\begin{align}
 & \frac{\partial^2}{\partial \tilde{t}^2}  \tilde{\phi} + 3 \frac{H}{m}\, \frac{\partial}{\partial
 \tilde{t}} \tilde{\phi}-\frac{\partial^2_{\tilde{\sbm{x}}}}{a^2}
 \tilde{\phi} + \tilde{V}_{\tilde{\phi}} = 0\,, \label{KG}
\end{align}
where we introduced the dimensionless coordinates $\tilde{t} \equiv mt$ and
$\tilde{\bm{x}} = m \bm{x}$, using the curvature of the potential around the minimum, $d^2 V/d \phi^2 \simeq m^2$. Here, $H$ is the Hubble parameter, $a$ is the scale factor and
$\tilde{V}_{\tilde{\phi}} \equiv d\tilde{V}/d\tilde{\phi}$. For a
power-law expansion with $a \propto t^p$, the Hubble parameter is given
by $H/m = p/\tilde{t}$. Since the equation of motion (\ref{KG}) is given by the dimensionless form, the dynamics of the axion does not depend on the values of $m$ and $f$. Therefore, as far as the assumptions described in the previous subsection are fulfilled, our discussion in this paper can apply to a spectator axion with a general parameter set $(m,\, f)$, including the QCD axion.~\footnote{When the axion is the dominant component of the Universe, the value of $H_{\rm osc}/m $ depends on $f/M_{\rm pl}$, where $M_{\rm pl}$ denotes the Planck mass. Therefore, the dynamics of the axion becomes rather different from the one for the spectator axion (a relevant discussion can be found e.g., in Ref.~\cite{Lozanov:2017hjm}). (For a more realistic study, the temperature dependence of the potential parameter should be taken into account.)}

For the homogeneous mode of $\tilde\phi$, the field equation (\ref{KG}) reads \footnote{In this paper, assuming $f$ is larger than the Hubble parameter during inflation and the symmetry restoration does not occur after inflation, we set the initial fluctuation of the axion to be a peturbatively small value. Meanwhile, a formation of axion clumps was discussed in the scenario where the symmetry breaking takes place after inflation e.g., in Refs.\cite{Kolb:1993hw,Kolb:1994fi,Turner:1985si,Vaquero:2018tib}.}

\begin{equation}
\frac{d^2}{d\tilde{t}^2}\,
 \tilde{\phi}+3\frac{H}{m}\frac{d}{d\tilde{t}}\, \tilde{\phi}+\tilde{V}_{\tilde{\phi}}=0\,.
 \label{eq:KGhomo}
\end{equation}
First, let us consider a quadratic potential with
$\tilde{V}=\tilde{\phi}^2/2$, which corresponds to the region
$|\tilde{\phi}| \ll 1$ in our setup. Then, the time evolution of
the homogeneous mode is divided into two stages. When the Hubble
friction dominates the potential driven force, $\phi$ undergoes a
slow-roll evolution, likewise an inflaton. On the other hand, when the
latter dominates the former, $\phi$ undergoes the harmonic oscillation
with the period $2\pi$ in $\tilde{t}$. For the quadratic potential, the transition
takes place at around $3 H_{\rm osc} \simeq m$. Here and hereafter, we put the index osc on
quantities evaluated at the onset of the oscillation. 

Meanwhile, when
$\tilde{V}$ for $|\tilde{\phi}| \geq 1$ is shallower than the quadratic
potential and the scalar field was initially located in this region,
$H_{{\rm osc}}/m$ becomes less than 1 or equivalently 
$\tilde{t}_{{\rm osc}}$ becomes larger than 1, indicating the significantly delayed onset of the oscillation compared with the quadratic potential case. As was argued in
Ref.~\cite{Kitajima:2018zco}, the reason for this can
be understood as follows. Equating the time scale of the cosmic
expansion with the one of the potential driven force, we can roughly
evaluate $H_{\rm osc}$ as 
\begin{equation}
 \frac{m}{H_{\rm osc}} \simeq \tilde{t}_{\rm osc} \simeq 
\sqrt{\left|\frac{\tilde{\phi}}{\tilde{V}_{\tilde{\phi}}}\right|_{\tilde\phi=\tilde\phi_i}}  \label{Exp:Hosc}
\end{equation}
or $H_{\rm osc} \simeq \sqrt{|(dV/d\phi)/\phi|_{\phi=\phi_i}}$ with $\phi_i$ ($\tilde\phi_i$) the initial amplitude of $\phi$ ($\tilde\phi$). In fact, we can
numerically confirm that when $\tilde\phi$ was initially located in a plateau region of $\tilde{V}$, $\tilde{t}_{\rm osc}$ can be roughly estimated by evaluating
the right hand side of Eq.~(\ref{Exp:Hosc}). As will be discussed later, the delayed onset of the oscillation leads to efficient and sustainable resonance instability. 
As a consequence, the instability can lead to a formation of oscillons or $i$-balls~\cite{Olle:2019kbo,Bogolyubsky:1976yu,Gleiser:1993pt,Copeland:1995fq,Gleiser:2009ys,Amin:2011hj,Kasuya:2002zs,Amin:2010jq,Lozanov:2019ylm,Ibe:2019vyo,Hong:2017ooe}, which are almost spherical clumps made up of the oscillating scalar field. Then, the empirically known condition for the delayed onset, i.e., the potential gradient $\tilde{V}_{\tilde{\phi}}$ is shallower than the one
for the quadratic potential, $\tilde{\phi}$, matches with the condition for the oscillon formation.

Next, we discuss the time evolution of the linear perturbation 
$\delta \tilde{\phi} \equiv \delta \phi/f$, whose equation of motion is
given by 
\begin{equation}
\frac{d^2}{d\tilde{t}^2}\delta\tilde{\phi}_k+3\frac{H}{m}\frac{d}{d\tilde{t}}\delta\tilde{\phi}_k+\left(\left(\frac{k}{am}\right)^2+\tilde{V}_{\tilde{\phi}\tilde{\phi}}\right)\delta\tilde{\phi}_k=0\,, 
\label{eq:KGinhomo}
\end{equation}
where $\delta\tilde{\phi}_k$ denotes the Fourier mode of
$\delta\tilde{\phi}$ and 
$\tilde{V}_{\tilde{\phi}\tilde{\phi}} \equiv d^2 \tilde{V}/ d \tilde{\phi}^2$. 
Here, we neglect the metric perturbations, which do not play an
important role in our present setup\footnote{The mode equation with the metric perturbation in Newtonian gauge is given by
\begin{equation}
 \frac{d^2}{d\tilde{t}^2}\delta\tilde{\phi}_k+3\frac{H}{m}\frac{d}{d\tilde{t}}\delta\tilde{\phi}_k+\left(\left(\frac{k}{am}\right)^2+\tilde{V}_{\tilde{\phi}\tilde{\phi}}\right)\delta\tilde{\phi}_k
  - 2 \tilde{V}_{\tilde{\phi}} \Phi_k + \frac{d \tilde{\phi}}{d \tilde{t}}
  \frac{d \Phi_k}{d \tilde{t}} =0\,, \label{Exp:pKGwmetric}
\end{equation}
where $\Phi_k$ denotes the Fourier mode of the curvature perturbation in this gauge. Noticing
that the amplitudes of $\tilde{V}_{\tilde{\phi} \tilde{\phi}}$,
$\tilde{V}_{\tilde{\phi}}$, $d \tilde{\phi}/d \tilde{t}$, and $\tilde{\phi}$
are roughly ${\cal O}(1)$ just after the onset of the oscillation, we
find that the last two terms in Eq.~(\ref{Exp:pKGwmetric}) are indeed
negligible for $k/aH \gg 1$. As will be discussed in the next
subsection, the resonance instability (in the dominant bands) takes place for the modes with
$k/(am) = {\cal O}(1)$, which are in sub Hubble scales for 
$H_{\rm osc}/m \ll 1$.}. Introducing $\varphi_k\equiv
a^{3/2}\delta\tilde{\phi}_k$, we can rewrite Eq.~(\ref{eq:KGinhomo}) as
\begin{equation}
\frac{d^2}{d\tilde{t}^2} \varphi_k (\tilde{t}\,) +\omega_k^2 (\tilde{t}\,) \varphi_k (\tilde{t}\,) =0\,, \label{Eq:fk}
\end{equation}
where $\omega_k$ denotes the (dimensionless) frequency given by
\begin{equation}
\omega_k^2=\left(\frac{k}{am}\right)^2+\tilde{V}_{\tilde{\phi}\tilde{\phi}}-\frac{3}{2}\frac{\dot{H}}{m^2}-\frac{9}{4}\left(\frac{H}{m}\right)^2,
\label{eq:omega_def}
\end{equation}
with $\dot{H}=dH/dt$.
In the next subsection, we will analyze the equation (\ref{Eq:fk}) in detail.

In this paper, we assume that the scalar field $\phi$ is not the
dominant component of the Universe. 
When $\phi$ is the dominant component, we also need to determine the geometry consistently by solving the Einstein equation.

\subsection{Resonance instability} 
When the background homogeneous mode starts to oscillate around the
potential minimum, the parametric resonance instability takes place, transferring the energy density of the homogeneous mode to the one of inhomogeneous modes
through the self-interaction of the scalar field. In this subsection, we overview the different types of the resonance instability, including the
flapping resonance instability~\cite{Kitajima:2018zco}. In particular, we focus on the case where the mode equation (\ref{Eq:fk}) can be written in the following form
\begin{align}
 & \frac{d^2 \varphi_k (\tilde{t}\,)}{d \tilde{t}^2} + \left[ A_k - 2 q \psi(\tilde{t}\,) \right] \varphi_k(\tilde{t}\,) = 0 \,,  \label{Eq:Hill}
\end{align} 
where $\psi(\tilde{t})$ is a quasi-periodic function, i.e., $\psi$ satisfies
 $\psi (\tilde{t} + T) \simeq \psi(\tilde{t}\,) $ at least in the time
 scale of the (background) oscillation. Here, we also assume that $A_k$ and $q$ stay
 almost constant in this time scale. This is indeed the case when the onset of the
 oscillation delays, taking $H_{\rm osc}/m \ll 1$.

When $\psi$ is exactly periodic, the above equation is called the Hill's
 equation. 
Note that the Hill's equation includes the well-known Mathieu equation where $\psi(\tilde{t}\,)$ is given by the sinusoidal function, i.e.,
\begin{align}
 &  \frac{d^2 \varphi_k(\tilde{t}\,)}{d \tilde{t}^2} + \left[ A_k - 2 q \cos 2
 \tilde{t}\, \right]  \varphi_k(\tilde{t}\,) = 0 \,. \label{Eq:Mathieu}
\end{align}

Here and hereafter, we define the quasi periodic function $\psi(\tilde{t}\,)$ so that $q$ becomes positive.
We normalize $\psi(\tilde{t}\,) $ by requesting $\langle \psi^2(\tilde{t}\,) \rangle =1/2$ in the same way as $\psi(\tilde{t}\,) =\cos 2\tilde{t}$, 
which satisfies $\langle \cos^2(2\tilde{t}\,) \rangle=1/2$. Here, we have introduced the square brackets to denote the time average over $T$, i.e., 
\begin{align}
 &  \langle F(\tilde{t}) \rangle \equiv \frac{1}{T} \int^{\tilde{t}+
 \frac{T}{2}}_{\tilde{t}- \frac{T}{2}} d \tilde{t}' F(\tilde{t}') \,, 
\end{align}
with $F(\tilde{t}\,)$ a general (quasi) periodic function of time.


\subsubsection{Parametric resonance for Mathieu equation} \label{sec:paramres}
Depending on the potential shape and the initial condition, different types of the parametric resonance instability become dominant. First, let us consider the case where the scalar potential $\tilde{V}$ can be expanded as
\begin{align}
 & \tilde{V}(\tilde{\phi})= \frac{\tilde{\phi}^2}{2} + \frac{\lambda}{4}\,
 \tilde{\phi}^4 + {\cal O} \left( \tilde{\phi}^6 \right)\,,  \label{Exp:Vpert}
\end{align}
with $\lambda<0$, keeping the self-interaction a small correction
to the quadratic potential. Let us emphasize that the perturbative expansion of $\tilde{V}$ is not assumed except for Sec.\ref{sec:paramres}. When the time scale of the cosmic expansion
is much longer than the one of the oscillation, we obtain an approximated solution of
the background homogeneous mode as
$\tilde{\phi}(\tilde{t}) \simeq \tilde{\phi}_{\ast}\cos\tilde{t}$. Then, the mode
equation (\ref{eq:omega_def}) is given by the Mathieu equation with 
\begin{equation}
  q=-\frac{3}{4}\lambda\tilde{\phi}_{\ast}^2\,, \qquad\qquad
  A_k=\left(\frac{k}{am}\right)^2- 2 q +1. \label{Exp:paraM}
\end{equation}

Now, let us briefly summarize the basic property of the Mathieu
equation. A more detailed explanation can be
found e.g., in Refs.~\cite{mclachlan1947theory, Kofman:1997yn}. The
Mathieu equation has the instability bands around $A_k \simeq n^2$ with
$n=1,\, 2,\, \cdots$, whose widths amount to
$(q/A_k)^n$\,. Therefore, as we increase the value of $q$, each instability band becomes wider, implying the resonance instability for a wider domain of the wave numbers. Because of that, the
parametric resonance with $q \gg 1$ is called the broad resonance and
the one with $q \ll 1$ is called the narrow resonance. For the narrow resonance,  the most
prominent instability takes place for the first resonance band with
$n=1$. The growth rate $\bar{\mu}$, with which $\varphi_k$ scales as $\varphi_k \propto e^{\bar{\mu} \tilde{t}}$, is given by $\bar{\mu} = q/2$ for the first band. This shows that a larger $q$ leads to more rapid exponential growth. For the broad resonance with $q \gg 1$, a significant particle production takes place at the moment when the adiabatic condition is strongly violated~\cite{Kofman:1997yn}.

Since the width of the resonance band is different, the shape of the
spectrum generated by the broad resonance is rather different from the one by
the narrow resonance. For the latter, the spectrum has sharp peaks at
the wavenumbers which correspond to $A_k \simeq n^2$. On the other hand,
for the former, the spectrum does not have a peak and all the modes
below the critical wavenumber $k_{\ast}$, i.e., $k\leq k_{\ast}$, are
enhanced.

In an expanding Universe, as the amplitude of the homogeneous field
decays due to the cosmic expansion, the parameter $q$ decreases in
time. As a result, the resonance instability becomes less and less
efficient. In particular, for $H_{\rm osc} \simeq m$, since the cosmic
expansion redshifts away the excited mode, the narrow resonance is
rather inefficient and does not continue sustainably. On the other hand,
as was discussed in Ref.~\cite{Soda:2017dsu}, when the oscillation takes
place much later with $H_{\rm osc} \ll m$, the cosmic expansion is no
longer important and the parametric resonance can continue (typically
until the backreaction becomes significant).

Here, we also discussed the parameter range with $q \gg 1$, where the
broad resonance takes place. However, this parameter choice contradicts with the assumption that the
self-interaction is perturbatively suppressed. In the next section, we
will argue that the broad resonance hardly takes place for the
self-resonance especially in an expanding Universe. Related to this, it
has been sometimes said that the self-resonance is typically
inefficient. However, in Sec.~\ref{Sec:Hill}, we will show that the flapping resonance provides an efficient self-resonance which can take place as well as in an expanding Universe.

\subsubsection{Flapping resonance} \label{sec:intermres}
The flapping resonance instability, pointed out in
Ref.~\cite{Kitajima:2018zco}, is qualitatively different from either of
the broad and narrow resonance instability. This instability occurs
during an oscillation which goes across the inflection points. During this period, since
$\tilde{V}_{\tilde{\phi} \tilde{\phi}}$ changes the signature,
$\omega^2_k$ also changes the signature for low-$k$ modes. (The flapping
resonance was named after the time evolution of the potential for $\delta \phi_k$ which visually reminds us of a flapping of wings.) Since
$\omega^2_k$ changes the signature, the adiabatic condition is
significantly violated when the flapping resonance takes place. Then, one may expect
that the flapping resonance is a special case of the broad
resonance. However, unlike the typical broad resonance, 
the flapping resonance generates a spectrum with a peak. The kinematic understanding
of the flapping resonance was discussed in
Ref.~\cite{Kitajima:2018zco}. 

\begin{figure}
\centering
\includegraphics[height=5cm, width=.48\textwidth]{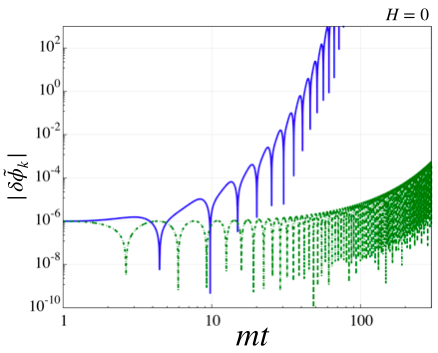}
\hfill
\includegraphics[height=5cm, width=.48\textwidth]{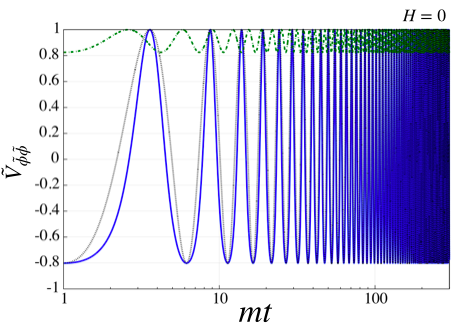}
\caption{The left panel shows the time evolution of $\delta \tilde{\phi}_k$ for the cosine potential. The green dotted line is for $\tilde{\phi}_i=0.60$ with $k/m$=1.0 and the blue line is for $\tilde{\phi}_i=2.5$ with $k/m$=0.6. Here, we chose the wavenumbers which undergo the resonance instabilities for these three cases.\label{Fg:evolution_typ} The right panel shows the time evolution of $\tilde{V}_{\tilde{\phi} \tilde{\phi}}$ for these two cases (blue solid, green dotted). For a comparison, we showed, by the black dotted line, the cosine function ($\propto \cos (2 \pi mt/T)$) whose amplitude and the mean value are adjusted to $\tilde{V}_{\tilde{\phi}\tilde{\phi}}$ for $\tilde{\phi}_i=2.5$.} 
\end {figure}

The left panel of Fig.~\ref{Fg:evolution_typ} shows the time evolution of $\delta \phi_k$ for the cosine potential under two different initial conditions for the homogeneous mode. Here, for an illustrative purpose, we neglected the cosmic expansion. The green dotted line shows the time evolution for $\tilde{\phi}_i=0.60$ and the blue line shows the one for $\tilde{\phi}_i=2.5$. The right panel of Fig.~\ref{Fg:evolution_typ} shows the time evolution of the second derivative of the scalar potential $\tilde{V}_{\tilde{\phi} \tilde{\phi}}$. For $\tilde{\phi}_i=2.5$, the signature of $\tilde{V}_{\tilde{\phi} \tilde{\phi}}$ periodically changes during the rapid growth. As a result, $\omega_k^2$ for the low-$k$ modes also changes the signature, leading to the significant violation of the adiabatic condition. Meanwhile, the adiabatic condition is well preserved for $\tilde{\phi}_i=0.60$. The resultant spectrums for these two cases both have peaks unlike a spectrum generated by a typical broad resonance instability. The peak width for $\tilde{\phi}_i=0.60$ is narrower than the one for $\tilde{\phi}_i=2.5$. These aspects lead us to conclude that the dominant instability for $\tilde{\phi}_i=2.5$ is the flapping resonance and the one for $\tilde{\phi}_i=0.60$ is the narrow resonance. 

Let us emphasize that when the flapping resonance is the dominant instability, the quasi periodic function $\psi(\tilde{t}\,)$ in $\omega^2_k$ is not the sinusoidal function. In fact, as shown in the right panel of Fig.~\ref{Fg:evolution_typ}, the time evolution of $\tilde{V}_{\tilde{\phi} \tilde{\phi}}$ for $\tilde{\phi}_i=2.5$ (blue solid) significantly deviates from the sinusoidal function (black dotted). 
This is because the flapping resonance takes place when the self-interaction is not perturbatively suppressed. As a result, the structure of the resonance band becomes rather different from the one for the Mathieu equation. This motivates us to analyze the stability/instability chart for the general Hill's equation in an expanding Universe more carefully, generalizing the analysis for the Mathieu equation.

\section{Resonance instability for Hill's equation}  \label{Sec:Hill}
When the self-interaction of the scalar field is not perturbatively suppressed, the quasi periodic function $\psi(\tilde{t}\,)$, in general, cannot be given by an analytic
function. Then, it is somewhat cumbersome to compute the value of $q$ in Eq.~(\ref{Eq:Hill})
after identifying the normalized quasi-periodic function $\psi(\tilde{t}\,)$. In this section, proposing a way to compute $q$ directly (without the identification of $\psi(\tilde{t}\,)$), we show that $q$ characterizes the parametric resonance instability for the general Hill's equation, including case where the oscillation of the homogeneous mode is anharmonic.

\subsection{New parameter for classification}  \label{SSec:classification}
\subsubsection{Definition}   
Compared to the one for the Mathieu equation, the structure of the
resonance band for the general Hill's equation has not been well
investigated especially in a cosmological context (yet see Refs.~\cite{Koutvitsky_PLA,Koutvitsky:2005ds,Koutvitsky:2015mda,Koutvitsky:2018afp,Braden:2010wd,Zanchin:1997gf,Adams:2009ioa,Lachapelle:2008sy}). In particular, a qualitative understanding of the stability/instability chart for the general Hill's equation has been still missing. 
In this section, we discuss a way
to characterize different types of the parametric resonance
instability described by the Hill's equation.

One may think that a possible reference for the classification is (the
degree of) the violation of the adiabatic condition. However, as we
argued in the previous section, the adiabatic condition is significantly
violated both for the broad resonance and the flapping resonance, while
the resultant spectrums are very different for these cases. 
Here, let us
recall that for the Mathieu equation, a large $q$ implies a significant
violation of the adiabatic condition and a large growth rate (inside the resonance band). However, the violation of the adiabatic condition does not directly imply a large growth rate. In fact, when the adiabatic condition is violated, we find
\begin{align}
 & \left| \frac{d \omega_k/ d \tilde{t}}{\omega^2_k} \right| 
=  \frac{q |d\psi(\tilde{t}\,)/d \tilde{t}|}{|A_k - 2 q
 \psi(\tilde{t}\,)|^{3/2}} \gg 1\,,   \label{Cond:Ad}
\end{align}  
and this inequality is fulfilled for $A_k \sim 2q$, even when the growth rate (nor $q$) is not particularly large. Therefore, the violation of the adiabatic condition does not necessarily ensure a significant growth of the fluctuation even for the Mathieu equation.

Having this lesson in mind, here we investigate whether the parameter $q$ in
Eq.~(\ref{Eq:Hill}), still characterizes the growth rate and the width
of the resonance band for the general Hill's equation or not. Using 
$\langle \psi^2 (\tilde{t}\,) \rangle = 1/2 $ and 
$\langle \psi(\tilde{t}\,) \rangle =0$, we can compute the parameter $q$ as 
\begin{equation}
q = \sqrt{\frac{\langle(\omega_k^2-\langle\omega_k^2\rangle
)^2\rangle}{2}} \, \,  \left( \equiv \tilde{q} \right). 
\label{eq:qdef}
\end{equation}
Notice that the middle expression can be computed without identifying
the quasi periodic function $\psi (\tilde{t}\,)$. We call thus
calculated quantity as $\tilde{q}$, distinguishing $q$ whose calculation
requires identification of $\psi(\tilde{t}\,)$. Since we used the
same normalization for $\psi(\tilde{t})$ as the one for
$\psi(\tilde{t})= \cos (2 \tilde{t}\,)$, when the mode equation for
$\varphi_k$ takes the form of the Mathieu equation, $\tilde{q}$ coincides with $q$ in the
Mathieu equation. When the background homogeneous mode has a significant
deviation from the harmonic oscillation, in general, we need a numerical computation
to evaluate $\tilde{q}$. When the frequency $\omega_k$ is given by
Eq.~(\ref{eq:omega_def}), we can further rewrite Eq.~(\ref{eq:qdef}) as
\begin{equation}
\tilde{q}=\sqrt{\frac{\langle(\tilde{V}_{\tilde{\phi}\tilde{\phi}}-\langle\tilde{V}_{\tilde{\phi}\tilde{\phi}}\rangle)^2\rangle}{2}},  
\end{equation}
where the non-oscillatory contributions dropped, being canceled between
the two terms. In this paper, having an axion in mind, we analyze
scalar potentials which fulfill the properties listed in
Sec.~\ref{SSec:notation}. However, the parameter $\tilde{q}$ can be used more generically to characterize the different types of the parametric resonance which is described by the Hill's equation (\ref{Eq:Hill}).


\subsubsection{Floquet analysis} 
To show that the new parameter $\tilde{q}$ indeed characterizes the width of the resonance band for the general Hill's equation, here we perform the Floquet analysis. The Floquet theorem states
that a general solution of the Hill's equation (\ref{Eq:Hill}) can be
expressed as
\begin{align}
 & \varphi_k(\tilde{t}\,) = P_1(\tilde{t}\,)\, e^{\mu_k \tilde{t}} + P_2(\tilde{t}\,)\,
 e^{- \mu_k \tilde{t}} \label{Exp:Hillsol}
\end{align}
where $P_i(\tilde{t}\,)$ with $i=1,\,2$ denote the periodic functions
which satisfy $P_i(\tilde{t}\,) = P_i(\tilde{t} +T)$ and $\mu_k$ is a
complex number, so called the Floquet exponent. 
When the real part of
$\mu_k$ is positive, $e^{\mu_k \tilde{t}}$ exponentially grows. A convenient algorithm to compute the Floquet exponent is summarized, e.g., in Ref.~\cite{Amin:2014eta}.

\begin{figure}
\centering
\includegraphics[height=6cm,width=.48\textwidth]{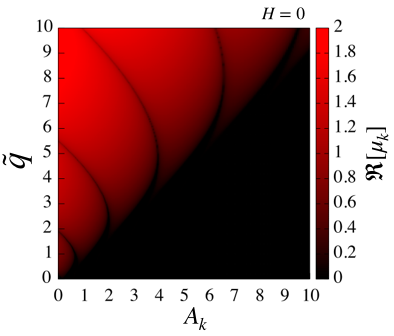}
\hfill
\includegraphics[height=6cm,width=.48\textwidth]{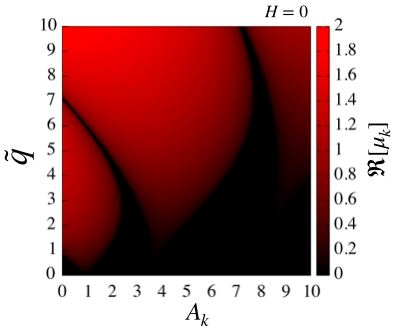}
\caption{These plots show the real part of the Floquet exponent, $\Re[\mu_k]$, for the Hill's equation. For the left panel, we used $\psi(\tilde{t}\,)$ computed for the cosine potential with $\tilde{\phi}_i =  2.5$ and for the right panel, we used the one with $\tilde{\phi}_i = 0.60$. For the right panel, since $\psi(\tilde{t}\,)$ is not very different from the sinusoidal function, the stability/instability chart is close to the one for the Mathieu equation. Notice that since we use the dimensionless time coordinate $\tilde{t} = mt$, the Floquet exponent $\mu_k$ is dimensionless. 
\label{Fg:Hill_Floquet}}
\end {figure}
To analyze the parametric resonance for a wide range of $\tilde{q}$, let us consider a toy model described by the Hill's equation whose frequency is given by 
$\omega_k^2 = A_k - 2 q \psi(\tilde{t}\,)$ with  
\begin{equation}
 \psi(\tilde{t}\,) \equiv \frac{\tilde{V}_{\tilde{\phi} \tilde{\phi}} (\tilde{t}\,) - \langle\tilde{V}_{\tilde{\phi} \tilde{\phi}} (\tilde{t}\,) \rangle }{\sqrt{2\, \langle(\tilde{V}_{\tilde{\phi} \tilde{\phi}} (\tilde{t}\,) - \langle\tilde{V}_{\tilde{\phi} \tilde{\phi}} (\tilde{t}\,) \rangle)^2 \rangle}}\,. \label{eq:psi}
\end{equation} 
With this normalization, $q$ in $\omega_k^2$ immediatelly gives $\tilde{q}$. In this toy model, we can choose an arbitrary value for the amplitude of the oscillatary term in $\omega_k^2$ by changing $q$. Here, we determine the periodic function $\psi(\tilde{t}\,)$ by using the time evolution of $\tilde{V}_{\tilde{\phi} \tilde{\phi}} (\tilde{t}\,)$ for the cosine potential with the initial condition $\tilde{\phi}_i = 0.60$ and 2.5. As shown in Fig.~\ref{Fg:evolution_typ}, the time evolution of $\tilde{V}_{\tilde{\phi} \tilde{\phi}} (\tilde{t}\,) = \cos \tilde{\phi} (\tilde{t}\,)$ significantly deviates from the harmonic oscillation for $\tilde{\phi}_i=2.5$. Notice that the Hill's equation with $\omega_k^2 = A_k - 2 q \psi(\tilde{t}\,)$ does not describe an explicit model of the self-resonance except for 
$- 2 q = \sqrt{2\, \langle(\tilde{V}_{\tilde{\phi} \tilde{\phi}} (\tilde{t}\,) - \langle\tilde{V}_{\tilde{\phi} \tilde{\phi}} (\tilde{t}\,) \rangle)^2 \rangle}$.

\begin{figure}
\centering
\includegraphics[width=7.5cm]{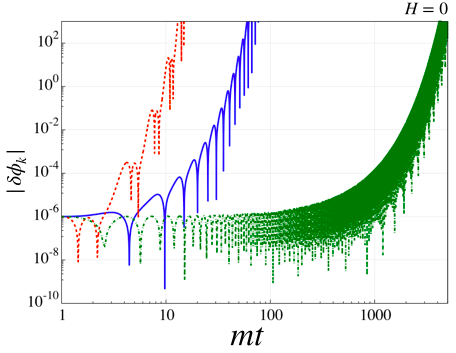}
\caption{This plot shows the time evolution of $\delta\tilde{\phi}_k (=\varphi_k)$ described by the Hill's equation with $\psi(\tilde{t}\,)$ determined by using $\tilde{V}_{\tilde{\phi} \tilde{\phi}} (\tilde{t}\,) = \cos \tilde{\phi} (\tilde{t}\,)$ for $\tilde{\phi}_i = 2.5$. The wavenumbers are picked up from the ones in the first resonance band. Here, we choose the three different values of $\tilde{q}$ as $\tilde{q}= 10$ with $k/m$=0.1(red), $\tilde{q}= 0.44$ with $k/m$=1.0 (blue), $\tilde{q}= 0.01$ with $k/m$=1.0 (green). The Hill's equation for $\tilde{q}= 0.44$ describes the self-resonance for the cosine potential with $\tilde{\phi}_i = 2.5$.
\label{Fg:Hill_evolution}}
\end {figure}
The left panel of Fig.~\ref{Fg:Hill_Floquet} shows the real part of the Floquet exponent, $\Re[\mu_k]$, for different values of $(A_k,\, \tilde{q})$ when we use $\tilde{V}_{\tilde{\phi} \tilde{\phi}}$ for $\tilde\phi_i=2.5$. Likewise, the stability/instability chart for the Mathieu equation, as we increase $\tilde{q}$ for a given $A_k$ in the resonance band, the band width becomes wider and the growth rate becomes larger \footnote{We also have conducted the Floquet analysis by using the periodic function $\psi(\tilde{t}\,)$  defined by Eq.(\ref{eq:psi}) where the potential is given by the broken power law potential (\ref{eq:power-law}) for the parameter sets $(c, \tilde{\phi}_i) =(-0.4,20),(-0.1,100),(0.1,100),(0.5,5)$. We find that for all of these examples, the growth rate $\Re[\mu]$ becomes larger and the band width becomes wider as we increase $\tilde{q}$ for a given $A_k$ in the resonance band.  } . For a comparison, in the right panel of Fig.~\ref{Fg:Hill_Floquet}, we show $\Re[\mu_k]$ for $\tilde{V}_{\tilde{\phi} \tilde{\phi}} (\tilde{t}\,)$ with $\tilde{\phi}_i = 0.60$. In this case, since $\psi(\tilde{t}\,)$ can be approximated by the sinusoidal function, the stability/instability chart becomes similar to the one for the Mathieu equation. These two plots in Fig.~\ref{Fg:Hill_Floquet} show that the stability/instability chart indeed becomes rather different due to the anharmonic oscillation. For $\tilde{\phi}_i = 0.60$, the resonance band extends to smaller $\tilde{q}$. This is because the periodic function $\psi(\tilde{t}\,)$ for $\tilde{\phi}_i = 2.50$ stays the negative region longer during each period of the oscillation than the one for $\tilde{\phi}_i = 0.60$. Figure \ref{Fg:Hill_evolution} shows the time evolution of $\delta \phi_k = \varphi_k$ (since there is no cosmic expansion)  for the three different values of $\tilde{q}$, when $\psi(\tilde{t}\,)$ is computed for $\tilde{\phi}_i = 2.50$ (the left panel of Fig.~\ref{Fg:Hill_Floquet}).   

\subsubsection{Self-resonance described by Hill's equation}
As we have discussed in this section, the parameter $\tilde{q}$ indeed characterizes the width of the resonance band for the parametric resonance described by the Hill's equation (\ref{Eq:Hill}). Therefore, using $\tilde{q}$, we categorize the parametric resonance as 
\begin{enumerate}
 \item $\tilde{q} \gg 1$: Broad resonance

 \item $\tilde{q} = {\cal O}(1)$: Intermediate resonance $\supset$ Flapping resonance

 \item $\tilde{q} \ll 1$: Narrow resonance
\end{enumerate}
For the broad and intermediate resonance, the adiabatic condition is strongly violated. Especially for the self-resonance, when $\tilde{q} \geq {\cal O} (1)$, the mode equation (\ref{Eq:fk}) takes a rather different form from the Mathieu equation. Meanwhile, for the narrow resonance, the adiabatic
condition is satisfied and the mode equation (\ref{Eq:fk}) is approximately given by
the Mathieu equation. In the succeeding subsections, we will show that
the flapping resonance~\cite{Kitajima:2018zco} is classified as the
intermediate resonance.

For all the potentials we studied, we did not find a prominent growth due to the broad resonance with $\tilde{q} \gg 1$ through the self-resonance. Instead, the flapping resonance is the most efficient
instability among these cases. This may be somewhat counter intuitive,
because when the mode equation is given by the Mathieu
equation, the growth rate becomes larger and larger as we increase $q$
(for a given $A_k$ in the resonance bands). We can naively understand the
reason as follows. First, let us consider $\tilde{V}$ whose curvature
$\tilde{V}_{\tilde{\phi} \tilde{\phi}}$ stays positive all the
time. Then, when we increase $\tilde{q}$ by considering a large self-interaction,
the non-oscillatory contribution $A_k$ also should increase, stabilizing
the system. Using $\omega^2_k$ for Hill's equation, this can be
understood by looking at 
$\omega^2_k \simeq A_k - 2 q \psi(\tilde{t}) \simeq \tilde{V}_{\tilde{\phi} \tilde{\phi}} > 0$ 
for the low-$k$
modes. (See Fig.~\ref{Fg:Hill_Floquet}, where the growth rate does not significantly increase, when we increase both $A_k$ and
$q$.)  Meanwhile, when we consider a scalar potential whose second derivative, $\tilde{V}_{\tilde{\phi} \tilde{\phi}}$, becomes negative during the oscillation, one may think that we can increase $\tilde{q}$ without enhancing $A_k$. However, when 
$- \tilde{V}_{\tilde{\phi} \tilde{\phi}}$ takes a large value,
$\tilde{V}$ becomes concave in this region, creating a bump. The concave
bump in $\tilde{V}$ disturbs the homogeneous mode to oscillate going
beyond this region. 
This becomes even more difficult in an expanding Universe, where the amplitude of the oscillation decreases due to the
Hubble friction.

Because of that, a prominent growth due to the broad resonance hardly
takes place through a self-interaction of a scalar field. 
Notice that a scalar potential which is connected to a plateau region for 
$|\tilde{\phi}| > 1$ such as the pure natural potential
(\ref{purenatural}) provides an optimized example where the potential
curvature takes negative values without creating a bump in $\tilde{V}$. This is the case where the inhomogeneous mode grows most significantly through the flapping resonance among the examples we addressed. In this paper, we only consider the resonance instability due to a self-resonance. By contrast, when we consider interactions between an oscillating scalar field and other fields, we can easily find examples where a prominent broad resonance with $\tilde{q} \gg 1$ takes place~(see, e.g., Refs.~\cite{Adshead:2018doq,Dufaux:2006ee}).

\subsection{Examples of self-resonance} 
In the previous subsection, using $\tilde{q}$, we introduced the classification of the resonance instability
for the Hill's equation into the three different types. In this subsection, we discuss a couple of potentials $\tilde{V}$ where the different resonance instability takes place. Here, including the cosmic expansion, we consider a scalar field which starts to oscillate in radiation dominated era like an axion dark matter.

\subsubsection{Scalar potentials}
\begin{figure}
\centering
\includegraphics[height=6cm,width=.48\textwidth]{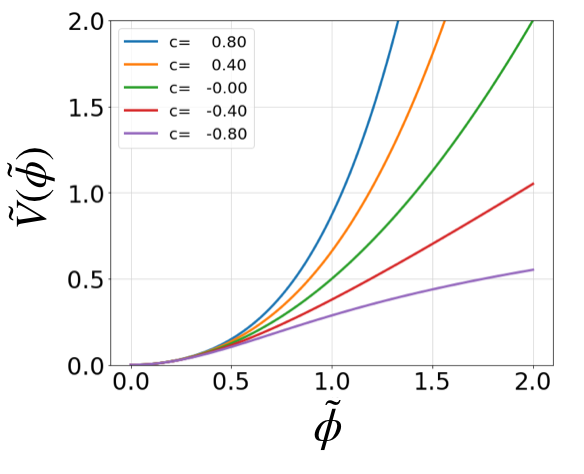}
\hfill
\includegraphics[height=6cm,width=.45\textwidth]{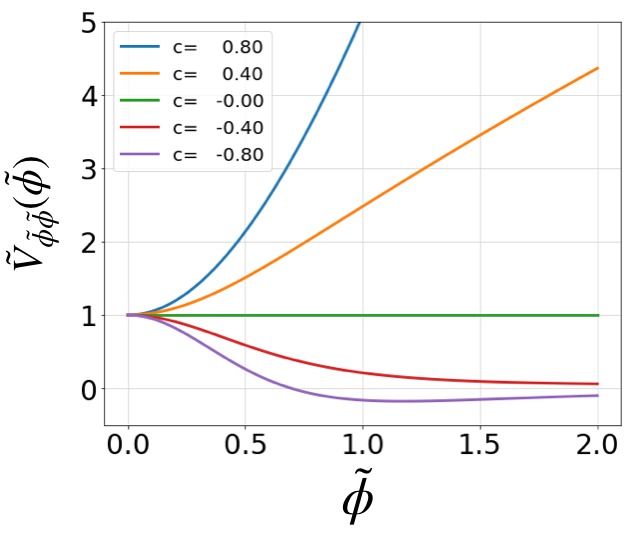}
\caption{\label{fig:model}The left panel shows the broken-power law
 potential and the right panel shows the curvature of the potential.}
\end{figure}
In this subsection, as examples, we discuss the broken power-law
potential, given by
\begin{equation}
\tilde{V}(\tilde{\phi})=\frac{1}{2}\tilde{\phi}^2(1+\tilde{\phi}^2)^c , \label{eq:power-law}
\end{equation}
and the scalar potential for the pure natural inflation, given in
Eq.~(\ref{purenatural}). Figure \ref{fig:model} shows the potential shape
(the left panel) and its second derivative (the right panel) for the
broken power-law potential with different values of $c$. Compared to the
quadratic potential, $\tilde{V}$ becomes steeper for $c > 0$ and
shallower for $c< 0$ in the region $|\tilde{\phi}| > 1$. Since the broken
power-law potential has a negative curvature region roughly for
$c < - 0.5$, having two inflection points, the flapping resonance can occur in this parameter range. Meanwhile, the pure natural
potential, which approaches to the plateau region for 
$|\tilde{\phi}| \gg 1$, always have two inflection points.

\subsubsection{Spectrums}  \label{SSSec:spectrum}
\begin{figure}
\centering
\includegraphics[width=8cm]{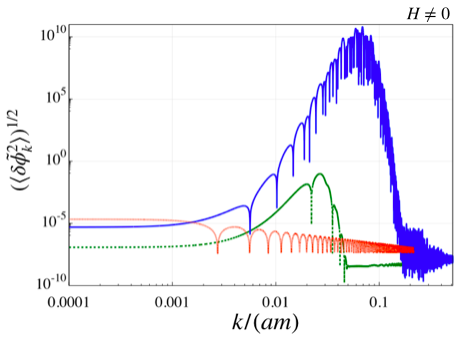}
\caption{This plot shows the spectrums for the
 broken power-law potential with $c=0.8$ (red dashed), $c=-0.1$ (green
 dotted), and $c=-0.8$ (blue), respectively. The spectrums for $c=0.8$
 and $c= -0.1$ are evaluated after $N=250$ oscillations of the
 homogeneous mode and the one for $c=-0.8$ is evaluated after $N=20$
 oscillations. Since the spectrum also oscillates, we evaluated the root mean square (time) average during each oscillation period. Here, we chose the initial amplitude of $\delta \phi_k$ as $10^{-6}$.}\label{fig:spectrum_all}
\end{figure}

Figure \ref{fig:spectrum_all} shows the spectrum of $\delta\tilde{\phi}_k$ for three different values of $c$. 
For $c=0.8$ (red dashed), the spectrum does not have any peak and all low-$k$ modes are enhanced,
indicating that $\tilde{q}$ takes a large value (while the growth rate is small since $A_k$ is also large). For
$c=-0.1$ (green dotted) and $c=-0.8$ (blue), the spectrums have the
narrow peak and the broader peaks. This indicates that the dominant instability for the
former is the narrow resonance and the one for the latter is the
intermediate resonance. Indeed, for $c=0.8$, $c=- 0.1$ and $c=- 0.8$,
$\tilde{q}$ amounts to $\tilde{q}=102$, $\tilde{q}=0.010$, and
$\tilde{q}=0.24$, respectively, explaining the different peak widths for these cases. Since $\tilde{V}_{\tilde{\phi} \tilde{\phi}}$ and $\omega^2_k$ (for low-$k$ modes) flip the signatures for $c= -0.8$ during the resonant growth, this
instability is driven by the flapping resonance. As we argued in Sec.~\ref{SSec:classification}, the flapping
resonance instability (for $c=- 0.8$) leads to the most rapid growth among these three examples. Therefore, we evaluated the spectrum and $\tilde{q}$ for $c=- 0.8$ only after $N=20$ time oscillations of the background
homogeneous mode, while we evaluated them for $c=0.8$ and $c=- 0.1$
after $N=250$ time oscillations. The damping of the homogeneous mode due to the
Hubble friction is the most prominent for $c=0.8$, since $\tilde{V}$ becomes steepest in $|\tilde{\phi}| > 1$, where the self-interaction does not vanish. Because of that, the resonant growth for $c=0.80$ terminates only after several oscillations without enhancing the inhomogeneous modes significantly. The physical wavenumber $k/(am)$ at the peak of the spectrum for $c= -0.10$ is
smaller than the one for $c= -0.80$. This is mainly because the redshift
due to the cosmic expansion was more influential for $c= -0.10$.

Here, we analyzed the spectrum of $\delta \phi_k$ based on the linear analysis. Once
the backreaction due to the enhanced inhomogeneous modes becomes important,
we cannot trust the result from the linear analysis. However, when the
dominant emission of the GWs takes place just after the inhomogeneous
mode becomes comparable to the homogeneous one, we expect that the
characteristic feature of the axion's spectrum computed from the linear
analysis can be inherited as well as in the spectrum of the emitted GWs. In that case, using the initial amplitude of the field fluctuation and the linear growth rate, we can estimate the emission time of the GWs by equating the amplitude of the inhomogeneous mode to the one of the background homogeneous mode.


\subsubsection{Floquet analysis}
Now, let us analyze the stability/instability chart for the broken power-law
potential and the pure natural potential, performing the Floquet analysis. In an expanding Universe, since the frequency $\omega_k$ is not a
periodic function, the mode equation (\ref{Eq:fk}) does not exactly take the
form of the Hill equation. Nevertheless, assuming that $\omega_k$ remains
periodic and the solution can be written as Eq.~(\ref{Exp:Hillsol}) in the time scale of the oscillation, we evaluate the growth rate $\Re[\mu_k]$. This assumption can be verified in particular for $H_{\rm osc}/m \ll 1$, where the inhomogeneity grows efficiently through the intermediate resonance or the narrow resonance. While we evaluate the Floquet exponent under the assumption that the solution is given in the form of Eq.~(\ref{Exp:Hillsol}), the cosmic expansion is properly taken into account in solving the mode equation.

\begin{figure}
\centering
\includegraphics[width=.48\textwidth]{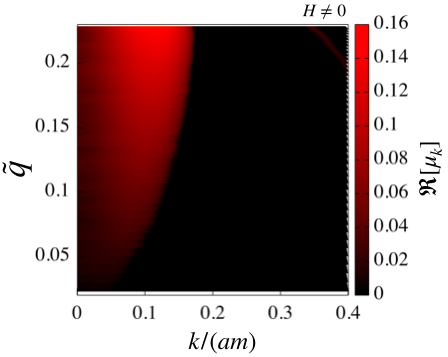}
\hfill
\includegraphics[width=.48\textwidth]{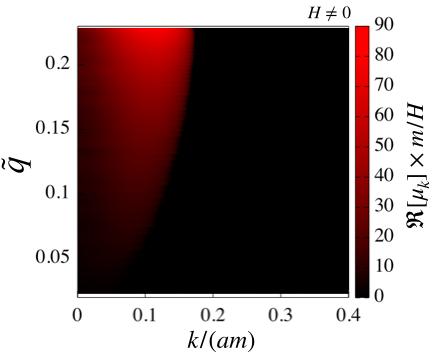}
\caption{These plots show the growth rate for the broken power-law
potential. Different colors show different values of the real part of the Floquet exponent, $\Re[\mu_k]$.}
\label{fig:inst_broken} 
\end{figure}

Figure \ref{fig:inst_broken} shows the growth rate $\Re[\mu_k]$
evaluated during the $N=20$ oscillation after the onset of the oscillation
for the broken power-law potential. When we change the model parameter $c$ ($-0.8 \leq c \leq -0.01$) 
for a given initial condition, since the background evolution changes, the parameter $\tilde{q}$ also varies. In Fig.~\ref{fig:inst_broken}, we used the obtained $\tilde{q}$ for the vertical axis. In this range, the dominant resonance instability is either the narrow resonance or the
intermediate resonance. The left panel of Fig.~\ref{fig:inst_broken} shows the growth rate given by the real part of the Floquet exponent. Notice that even if
the growth rate  $\Re[\mu_k]$, normalized by the mass scale $m$, is 
smaller than 1, the mode $k$ can grow exponentially in an expanding
Universe, as far as  
\begin{align}
 &  \Re[\mu_k]\, \frac{m}{H} \gg 1\,.  \label{eq:Floquetexp}
\end{align}
The right panel of Fig.~\ref{fig:inst_broken} shows that $\Re[\mu_k] \times (m/H)$
becomes much larger than 1 in a wide range of $\tilde{q}$, being enhanced by $m/H \gg 1$ (Taking the large value of $m/H$ indeed verifies computing the growth rate using the Floquet analysis.). For models where the flapping resonance takes place, the onset of the oscillation typically delays, taking $H_{\rm osc}/m \ll 1$. This is another reason why the flapping resonance typically becomes the most efficient self-resonance instability in an expanding Universe. (See also the discussion in the next subsection.)

Due to the anharmonic oscillation and the cosmic expansion, the
structure of the resonance band in the left panel of
Fig.~\ref{fig:inst_broken} is different from the one for the Mathieu
equation. However, several properties remain
unchanged. First, as was discussed in Sec.~\ref{SSSec:spectrum}, the
width of the resonance band becomes wider for a larger value of
$\tilde{q}$. Second, the growth rate $\Re[\mu_k]$ becomes larger, as we
increase $\tilde{q}$ for a given $k/(am)$. Meanwhile, for the broken power-law potential, the resonance band extends to a smaller value of $\tilde{q}$, compared to
the one for the Mathieu equation. This is mainly because as a
consequence of the anharmonic time evolution, 
$\omega^2_k$ stays longer in the negative region for the broken power-law potential than the case where the mode equation is given by the Mathieu equation.

\begin{figure}
\centering
\includegraphics[width=.48\textwidth]{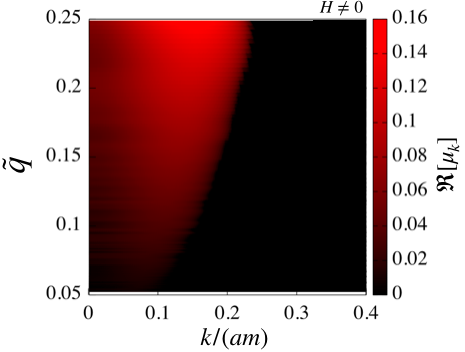}
\hfill
\includegraphics[width=.48\textwidth]{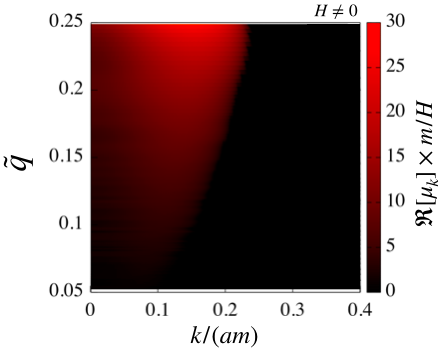}
\caption{These plots show $\Re[\mu_k]$ for the pure natural potential
 (\ref{purenatural}). }  
\label{fig:inst_purenatural}
\end{figure}

Figure \ref{fig:inst_purenatural} shows the growth rate $\Re[\mu_k]$
evaluated during the $N=6$ oscillation after the onset of the oscillation
for the pure natural potential. The structure of the resonance band is
similar to the one for the broken power-law potential, while there is a
slight model dependence.

\subsection{Duration of flapping resonance instability} \label{SSec:duration}
As we have discussed in this section, the flapping
resonance, categorized in the intermediate resonance, is typically the
most efficient resonance instability among the self-resonance instability. In this subsection, we investigate
when a sustainable flapping resonance takes place or more specifically
which quantity characterizes the duration of the flapping resonance.

In Sec.~\ref{SSec:classification}, we argued that when the flapping resonance takes place, the parameter $\tilde{q}$ amounts to ${\cal O}(1)$. In an
expanding Universe, $\tilde{q}$ decreases in time, since the
self-interaction becomes more and more suppressed, as the Hubble friction reduces the
amplitude of $\tilde{\phi}$. Therefore, one necessary condition
for a long-lasting intermediate resonance, including the flapping
resonance, is maintaining $\tilde{q}= {\cal O}(1)$ sufficiently long. In an
expanding Universe, this is possible when the onset of the oscillation
delays, i.e., $H_{\rm osc}/m \ll 1$.  

\begin{figure}
\centering
\includegraphics[height=5.5cm,width=.48\textwidth]{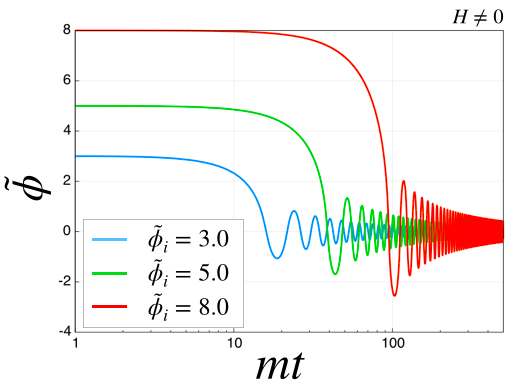}
\hfill
\includegraphics[height=5.5cm,width=.48\textwidth]{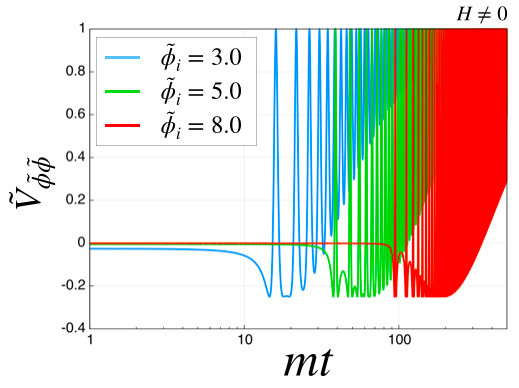}
\caption{\label{fig:duration}The left panel shows the time evolution of
 the background homogeneous mode and the right panel shows the time
 evolution of the curvature of the potential for the broken power-law
 potential with $c=-1.0$. Here, we set the initial value of the
 homogeneous mode to $\tilde{\phi}_i = 3.0, 5.0, 8.0$.}
\end{figure}

\begin{figure}
\centering
\includegraphics[width=8cm]{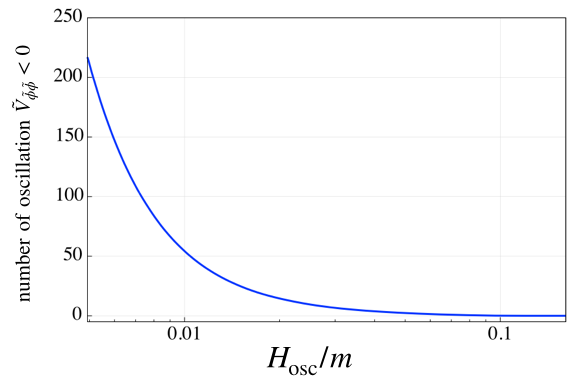}
\caption{This plot shows the relation between $H_{{\rm osc}}/m$ and
 the number of the oscillation during which $\tilde{V}_{\tilde{\phi} \tilde{\phi}}$ becomes negative. Here, we consider the broken power-law potential with $c = -0.8$ and change the initial value of the homogeneous mode $\tilde{\phi}_i \ (0.01\leq \tilde{\phi}_i \leq 50.0)$. }  
\label{fig:duration_H}
\end{figure}

Figure \ref{fig:duration} shows the time evolution of the
homogeneous mode $\tilde{\phi}$ and the curvature of the scalar
potential $\tilde{V}_{\tilde{\phi} \tilde{\phi}}$ for the broken
power-law potential. The left panel shows that for a larger value of
$|\tilde{\phi}_i|$, $H_{\rm osc}/m$ becomes smaller, making the cosmic
expansion during the oscillation less important. The right panel shows
that for a larger value of $|\tilde{\phi}_i|$, $\tilde{V}_{\tilde{\phi} \tilde{\phi}}$ indeed reaches the negative
region, going across the inflection points, during a larger number of the
oscillation. In this period, $\tilde{q}$ amounts to ${\cal O}(1)$
and the flapping resonance takes place. Figure \ref{fig:duration_H} shows more
explicitly the relation between $H_{\rm osc}/m$ and the number of the
oscillations during which $\tilde{V}_{\tilde{\phi} \tilde{\phi}}$ becomes 
negative. 

\begin{figure}
\centering
\includegraphics[height=5.5cm,width=.48\textwidth]{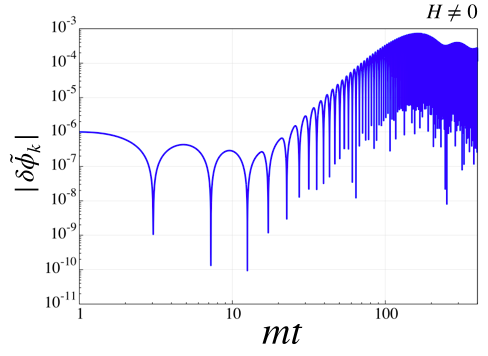}
\hfill
\includegraphics[height=5.4cm,width=.48\textwidth]{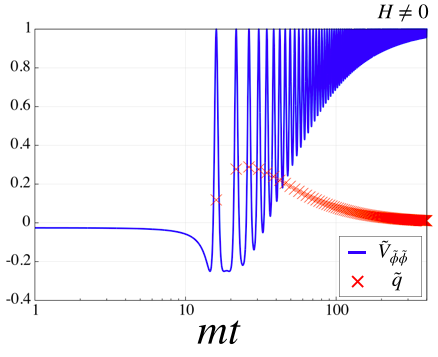}
\caption{\label{fig:notdelay}The left panel shows the time evolution of the inhomogeneous mode $\delta \tilde{\phi}_k$  with $k/(a_{{\rm osc}}m)=0.3$ and the right panel shows the time evolution of $\tilde{V}_{\tilde{\phi} \tilde{\phi}}$ and $\tilde{q}$ evaluated during each period of the oscillation. Here, we consider the broken power-law potential with $c=-1.0$ and $\tilde{\phi}_i=3.0$. } 
\end{figure}

When the cosmic expansion is not negligible at the onset of the
oscillation, the parameter $\tilde{q}$ gradually decreases in time as
the amplitude of the oscillating background homogeneous mode $\tilde{\phi}$ decreases. 
In this case, even if the intermediate resonance takes place just after $\tilde{\phi}$
commences the oscillation, the dominant instability turns into from the intermediate resonance to the narrow resonance. Figure \ref{fig:notdelay} shows one example where the
transition from the intermediate resonance to the narrow resonance takes
place around $\tilde{t}\, (=mt) \simeq 70$.

In Ref.~\cite{Soda:2017dsu}, it was shown that when the scalar potential of
an axion has a plateau region, e.g., as in the pure natural potential
(\ref{purenatural}) for $|\tilde{\phi}| > \sqrt{c}$, and the axion was
initially located at the plateau region, the onset of the oscillation
delays significantly, taking $H_{\rm osc}/m \ll 1$. Then, the flapping
resonance continues, maintaining $\tilde{q} = {\cal O}(1)$, without being
disturbed by the cosmic expansion until the inhomogeneous mode becomes
comparable to the homogeneous mode. The sustainable flapping resonance
leads to a copious emission of the GWs and the succeeding oscillon
formation~\cite{Kitajima:2018zco}.

\section{Does the flapping resonance persist for cosine potential?\label{sec:cos}}
In the previous section, we showed that a persistent flapping resonance can change an almost homogeneous distribution to a highly inhomogeneous one. The flapping self-resonance requires the potential region where the curvature is negative. The curvature for the cosine potential (\ref{Exp:cosine}), which has been conventionally considered for axions, is given by $\tilde{V}_{\tilde{\phi} \tilde{\phi}} = \cos \tilde{\phi}$, becoming negative around the potential maximum. Indeed, as shown in Sec.~\ref{Sec:Resonance}., the flapping self-resonance also takes place for the cosine potential, when we neglect the cosmic expansion. In this section, let us ask the question, "whether the flapping resonance persistently takes place for the cosine potential (\ref{Exp:cosine}) in an expanding Universe or not?"

In the previous section, we showed that the flapping resonance continues without being disturbed by the cosmic expansion, when $H_{{\rm osc}}/m$ is much smaller than 1. Therefore, we can rephrase our questions as whether $H_{{\rm osc}}/m$ can be sufficiently small or not for the cosine potential. Since the dimensionless cosine potential (\ref{Exp:cosine}) does not include any parameters, we investigate the possible values of $H_{{\rm osc}}/m$ by changing the initial condition. In particular, we change the initial velocity $\dot{\tilde{\phi}}_i \equiv d \tilde{\phi}/d (mt)|_{t=t_i}$, placing $\tilde{\phi}_i$ at the potential maximum, i.e., $\tilde{\phi}_i = \pi$.

\begin{figure}
\centering
\includegraphics[width=.47\textwidth]{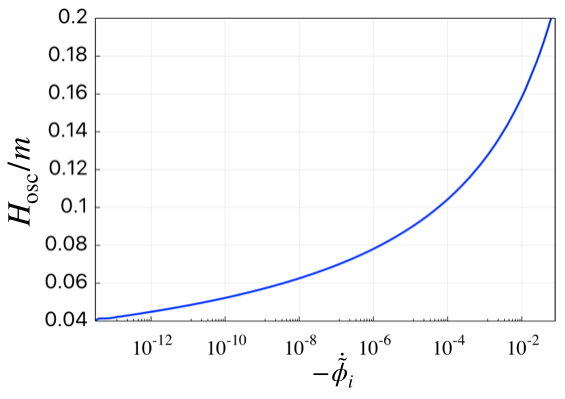}
\hfill
\includegraphics[width=.44\textwidth]{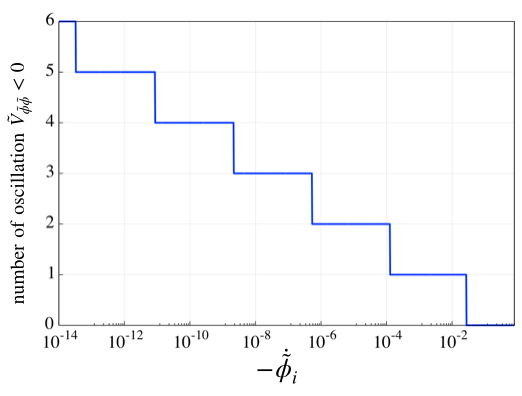}
\caption{The left panel shows the relation between the initial velocity $\dot{\tilde{\phi}}_i \equiv d \tilde{\phi}/d (mt)|_{t=t_i}$ and $H_{{\rm osc}}/m$ for the cosine potential. The vertical axis shows $H_{{\rm osc}}/m$ and horizontal axis shows the initial velocity of the homogeneous mode. We set the initial field value as $\tilde{\phi}_i=\pi$. The right panel shows the number of the oscillation during which the curvature of the potential becomes negative.}
\label{fig:cos_Hosc}
\end{figure}

The left panel of Fig.~\ref{fig:cos_Hosc} shows $H_{{\rm osc}}/m$ for different values of the initial velocity. Here, we again have considered the radiation dominated Universe with $a \propto t^{1/2}$. As is shown, even if we tune the initial velocity as $|\dot{\tilde{\phi}}_i| \sim 10^{-10}$, the onset of the oscillation does not delay significantly~\footnote{When we neglect the Hubble friction, we can understand the reason for the weak dependence on the initial velocity as follows. Indeed, for $|\dot{\tilde{\phi}}_i| \ll 1$, the Hubble friction is not very important except for the transition period between the slow-roll evolution and the oscillation. Solving the Klein-Gordon equation, we obtain the approximate solution of the homogeneous mode around the potential maximum as  
\begin{equation}
\tilde{\phi}(\tilde{t})=\pi+\frac{\dot{\tilde{\phi_i}}}{2}(e^{\tilde{t}}-e^{-\tilde{t}}).  \label{Exp:solcos}
\end{equation}
Here, we used $\tilde{V}_{\tilde{\phi}} \simeq - (\tilde{\phi}- \pi)$ around $|\tilde{\phi}- \pi| \ll 1$. Sinc ethe kinetic energy and the potential energy are roughly comparable at the onset of the oscillation, as a crude estimation, we use $|d \tilde{\phi}/d \tilde{t} (t_{\rm osc})| \sim |V(\tilde{\phi}(t_{\rm osc}))|  \sim O(1)$. Then, using Eq.~(\ref{Exp:solcos}), we obtain $\tilde{t}_{{\rm osc}} (= p\, m/H_{\rm osc}) =\cosh^{-1}(O(1)/|\dot{\tilde{\phi_i}}|)$, i.e., $H_{\rm osc}/m$ depends on the initial velocity only logarithmically.}. 
 Because of that, as shown in the right panel of Fig.~\ref{fig:cos_Hosc},even if the initial velocity is $|\dot{\tilde{\phi}}_i| \sim 10^{-10}$, the flapping resonance continues only 4 oscillation periods. Therefore, the long-lasting flapping resonance does not take place for the cosine potential, unless we extremely tune the initial condition.

\begin{figure}
\centering
\includegraphics[width=8.0cm]{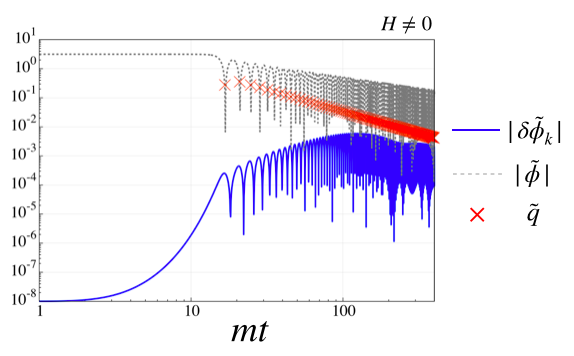}
\caption{This plot shows the time evolution of the inhomogeneous mode $\delta\tilde{\phi}_k$, the background homogeneous mode $\tilde{\phi}$, and $\tilde{q}$ evaluated during each period of the oscillation. The initial condition is set as $\tilde{\phi_i}=\pi$, $\dot{\tilde{\phi_i}}=-10^{-8}$ and we choose the wavenumber $k/m$=0.2.}
\label{fig:cos_bg}
\end{figure}

Figure \ref{fig:cos_bg} shows the time evolution of the background homogeneous mode (grey), the inhomogeneous mode with $k/(a_{{\rm osc}}m)=0.2$ (blue), and $\tilde{q}$. The low-$k$ modes are enhanced by the tachyonic instability before the onset of the oscillation. Since $\tilde{q}$ gradually decreases due to the Hubble friction during the oscillation, the flapping resonance turns to the narrow resonance.   

\begin{figure}
\centering
\includegraphics[width=15.0cm]{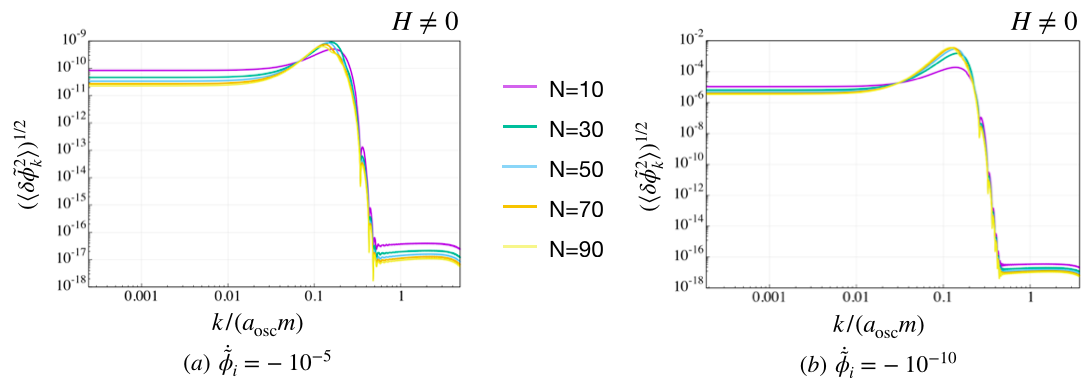}
\caption{\label{fig:cos_spectrum} The left panel shows the spectrum for $\dot{\tilde{\phi}}_i=- 10^{-5}$ and the right panel shows the one for $\dot{\tilde{\phi}}_i=- 10^{-10}$. The spectrums at different oscillation periods are shown by different colors. The wavenumber in the horizontal axis is divided by the scalar factor at the onset of the oscillation (and $m$).}
\end{figure}

The spectrum of the inhomogeneous mode $\delta\tilde{\phi}_k$ is shown in Fig.~\ref{fig:cos_spectrum}. The initial velocity is set to $\dot{\tilde{\phi}}_i=-10^{-5}$ in the left panel and $\dot{\tilde{\phi}}_i=-10^{-10}$ in the right panel. Here, the initial amplitude of the inhomogeneous mode is set to $10^{-15}$ so that the perturbed variables remain much smaller than their background values. 
For a smaller $|\dot{\tilde{\phi}}_i|$, the onset of the oscillation delays more, leading to a more efficient parametric resonance instability. The dominant instability in these cases is the narrow resonance, whose growth rate is much smaller than the one for the flapping resonance.

\section{Conclusion\label{sec:conc}}
In this paper, conducting the Floquet analysis for the Hill's equation, we showed that the growth rate and the width of the resonance band are characterized by the explicitly calculable parameter $\tilde{q}$, which measures the amplitude of the oscillatory contribution in $\omega_k^2$ likewise the parametric resonance described by the Mathieu equation. We also showed that the flapping resonance, found in Ref.~\cite{Kitajima:2018zco}, takes place for $\tilde{q}= O(1)$. We found that the flapping resonance typically leads to the most prominent growth of the inhomogeneity among the self-resonance instability, since the broad resonance with $\tilde{q} \gg 1$ hardly takes place, especially in an expanding Universe. We also showed that the flapping resonance does not persistently continue for the cosine potential, unless we extremely fine-tune the initial condition. Note that, if the potential is given by a plateau type potential like the purenatural model instead of the cosine potential and when the scalar field was initially located at an extensive potential region which is much shallower than the quadratic potential, the flapping resonance continues until the backreaction becomes important for a wide range of the initial velocity, because the initial velocity immediately decays.

In this paper, to establish a qualitative understanding of the resonance instability for the general Hill's equation in an expanding Universe, we focused on linear analysis. Obviously, this analysis cannot be applied, once the non-linear dynamics becomes important. Since the parametric resonance continues until the non-linear effects become important when the onset of the oscillation significantly delays, we need a numerical simulation such as the lattice simulation to understand the whole dynamics. Since the spectrum of the linear perturbation becomes rather different depending on the value of $\tilde{q}$, it is natural to expect that the succeeding dynamics will also depend on $\tilde{q}$. We will report this study in our forthcoming paper.

When the axion is the dominant component of dark matter, the parametric resonance of the axion dark matter can leave observable impacts in the spectrum of dark matter, which has been measured through various cosmological observations. To provide a theoretical prediction of the imprints, we need to include the metric perturbations and also other components of the Universe, using a realistic transfer function. We will leave this for a future study.


\acknowledgments
We would like to thank M.~Amin, K.~Lozanov and J.~Soda for helpful
discussions. We would like to thank Yukawa Institute for Theoretical Physics at Kyoto University, where this work was completed during the YITP-T-19-02 on "Resonant instabilities in cosmology". N.~K and Y.~U are supported by Grant-in-Aid for Scientific Research (B) under Contract No. 19H01894.
N.~K. acknowledges the support by Grant-in-Aid for JSPS
Fellows. Y.~U. is supported by JSPS Grant-in-Aid for Research Activity Start-up
under Contract No.~26887018, Grant-in-Aid for Young Scientists (B) under
Contract No.~16K17689, Grant-in-Aid for Scientific Research on
Innovative Areas under Contract Nos.~16H01095 and 18H04349, and the Deutsche Forschungsgemeinschaft (DFG,
German Research Foundation) - Project number 315477589 - TRR 211. Y.~U. is also supported in part by Building of Consortia for the Development of Human Resources in Science and Technology and Daiko Foundation.

\bibliographystyle{JHEP}

\end{document}